\documentstyle[jeep,graphics,equations,12pt,doublespace]{article}

\begin{document}           
\newpage

\begin{center}
 
{\Large  {\bf Waiting time to (and duration of) parapatric speciation}}\\

\vspace{.25in}

{\large Sergey Gavrilets} \\  

\end{center}

\vspace{.25in}
{\em Departments of Ecology and Evolutionary Biology 
and Mathematics,  
University of Tennessee, Knoxville, TN 37996 USA}

phone (865) 974-8136;
fax (865) 974-3067;
e-mail: sergey@tiem.utk.edu\\


Using a weak migration and weak mutation
approximation, I study the average waiting time to and the average duration
of parapatric speciation. The description of reproductive isolation used is
based on the classical Dobzhansky model and its recently proposed multilocus generalizations. The dynamics of parapatric speciation is modeled as a biased random walk with absorption performed by the average genetic distance between
the residents and immigrants. If a small number of genetic changes is 
sufficient for complete reproductive isolation, mutation and random genetic 
drift alone can cause speciation on the time scale of 10-1000 times the 
inverse of the mutation rate. Even relatively weak selection for local 
adaptation can dramatically decrease the waiting time to speciation.
The duration of parapatric speciation is shorter by orders 
of magnitude than the waiting time to speciation. For a wide range of 
parameter values, the duration of speciation is order one over the mutation 
rate.  In general, parapatric speciation is expected to be triggered by 
changes in the environment.\\


\section{1. INTRODUCTION}

Parapatric speciation is usually defined as the process of species
formation in the presence of some gene flow between diverging populations. 
From a theoretical point of view, parapatric speciation represents the most 
general scenario of speciation which includes both allopatric and
sympatric speciation as extreme cases (of zero gene flow and a very
large gene flow, respectively). 
The geographic structure of most species, which are usually composed of many
local populations experiencing little genetic contact for long periods of
time (Avise 2000), fits the one implied in the parapatric speciation scenario.
In spite of this, parapatric speciation has 
received relatively little attention compared to a large number of empirical 
and theoretical studies devoted to allopatric and sympatric modes 
(but see Ripley \& Beehler 1990; Burger 1995; Friesen \& Anderson 1997; 
Rol\'{a}n-Alvarez {\em et al.} 1997; Frias \& Atria 1998; Macnair \& Gardner 
1998). Traditionally, 
studies of parapatric speciation emphasized the importance of strong 
selection for local adaptation in overcoming the homogenizing effects
of migration (e.g. Endler 1977; Slatkin 1982). Recently, it has been 
shown theoretically that rapid parapatric speciation is possible even 
without selection for local adaptation if there are many loci affecting 
reproductive isolation and mutation is not too small relative to migration 
(Gavrilets {\em et al.} 1998, 2000{\em a}; Gavrilets 1999).

Earlier theoretical studies of speciation mostly concentrated on the 
accumulation of genetic differences that could eventually lead to complete 
reproductive isolation. However, within the modeling frameworks previously 
used complete reproductive isolation was not possible (but see Nei {\em et al.}
1983; Wu 1985).
Recently new approaches describing the whole process of speciation from origination to completion have been developed and applied to allopatric 
(Orr 1995; Orr \& Orr 1996; Gavrilets \& Hastings 1996; Gavrilets \& 
Boake 1998; Gavrilets 1999), parapatric (Gavrilets 1999; Gavrilets {\em et al.} 
1998, 2000a) and sympatric (e.g. Turner \& Burrows 1995; Gavrilets \& Boake
1998; van Doorn {\em et al.} 1998; Dieckmann \& Doebeli 1999) scenarios.
Here, I develop a new stochastic approach to modeling speciation as a biased
random walk with absorption. I use this framework to find the average waiting 
time to and the average duration of parapatric speciation. My results provides 
insights into a number of important evolutionary questions about the role 
of different factors (such as mutation, migration, random genetic drift, 
selection for local adaptation, genetic architecture of reproductive 
isolation) in controlling the time scale of parapatric speciation.

The method for modeling reproductive isolation adapted below is based on 
the classical Dobzhansky model (Dobzhansky 1937) discussed in detail in a 
number of recent publications (e.g. Orr 1995; Orr \& Orr 1996; Gavrilets \& 
Hastings 1996; Gavrilets 1997). The Dobzhansky model as originally described 
has two important and somewhat independent features (Orr 1995). First, the Dobzhansky model suggests that in some cases reproductive isolation can be reduced to interactions of ``complementary'' genes (that is genes that 
decrease fitness when present simultaneously in an organism). Second,
it postulates the existence of a ``ridge'' of well-fit genotypes that connects 
two reproductively isolated genotypes in genotype space. This ``ridge'' makes 
it possible for a population to evolve from one state to a reproductively isolated state without passing through any maladaptive states (``adaptive 
valleys''). 
The original Dobzhansky model was formulated for the two-locus case. The 
development of multilocus generalizations has proceeded in two directions.
A mathematical theory of the build up of incompatible genes leading to
hybrid sterility or inviability was developed by Orr (1995; Orr \& Orr 1996)
who applied it to allopatric speciation.
A complementary approach placing the most emphasis on ``ridges'' rather than
on ``incompatibilities'' was advanced by Gavrilets (1997, 1999; Gavrilets \& 
Gravner 1997; Gavrilets {\em et al.} 1998, 2000{\em ab}). This approach makes use 
of a recent discovery that the existence of  ``ridges'' is a general feature 
of multidimensional adaptive landscapes rather than a property of a specific genetic architecture (Gavrilets \& Gravner 1997; Gavrilets 1997, 2000). 
Here, I will use the ``ridges-based'' approach assuming that mating and the 
development of viable and fertile offspring is possible only between the organisms that are not too different over a specific set of loci
responsible for reproductive isolation. The adaptive landscape arising in 
this model is an example of ``holey adaptive landscapes'' (Gavrilets \& 
Gravner 1997; Gavrilets 1997, 2000) of which the original two-locus 
two-allele Dobzhansky model is the simplest partial case. 
My general results are directly applicable to the original Dobzhansky model.\\

\section{2. MODEL}

I consider a finite population of sexual diploid organisms with discrete 
non-overlapping generations. The population is subject to immigration 
from another population. For example, one can think of a peripheral 
population (or an island) receiving immigrants from a central population 
(or the mainland). All immigrants are homozygous and have a fixed
``ancestral'' genotype. Mutation supplies new genes in the population some 
of which may be fixed by random genetic drift and/or selection for local 
adaptation. Migration brings ancestral genes which, if fixed, will 
decrease genetic differentiation of the population from its ancestral 
state.

In this paper, I consider only the loci potentially affecting reproductive 
isolation. The degree of reproductive isolation depends on the extent of 
genetic divergence at these loci. Let $d$ be the number of loci at which 
two individuals differ. I posit 
that the probability, $w$, that two individuals are able to mate and
produce viable and fertile offspring is a non-increasing function of $d$ 
such that $w(0)=1$ and $w(d)=0$ for all $d>K$ where $K$ is a parameter of 
the model specifying the genetic architecture of reproductive isolation. 
This implies that individuals with identical genotypes at the loci under 
consideration are completely compatible whereas individuals that differ in 
more than $K$ loci are completely reproductively isolated. 
A small $K$ means that a small number of genetic changes is sufficient for
completely reproductive isolation. A large $K$ means that a significant 
genetic divergence is necessary for completely reproductive isolation. 
If $K$ is equal to the overall number of loci, complete reproductive 
isolation is impossible (neutral case). This simple 
model is appropriate for a variety of isolating barriers including 
premating, postmating prezygotic, and postzygotic (Gavrilets {\em et al.}
1998, 2000{\em ab}, Gavrilets 1999). I will allow the loci responsible
for reproductive isolation to have pleiotropic effects on the degree of
adaptation to the local environment (Gavrilets 1999; cf. Slatkin 1981; 
Rice 1984; Rice \& Salt 1988). Specifically, I will assume that each new 
allele potentially has a selective advantage $s$ ($\geq 0$) over the 
corresponding ancestral allele in the local environment.

I will use a weak mutation and weak migration approximation (e.g. Slatkin 
1976, 1981; Lande 1979, 1985; Tachida \& Iizuka, 1991; Barton 1993) 
neglecting within-population variation. Under this approximation the only 
role of mutation and migration is to introduce new alleles which quickly 
get fixed or lost. I will assume that the processes of fixation and loss of 
alleles at different loci are independent. Within this approximation, the 
relevant dynamic variable is the number of loci, $D_b$, at which a typical individual in the population is different from the immigrants. Variable
$D_b$ is the average genetic distance between residents and immigrants 
computed over the loci underlying reproductive isolation. 
The dynamics of speciation will be modeled as a 
random walk performed by $D_b$ on a set of integers $0,1,\dots,K,K+1$. In 
what follows I will use $\lambda_i$ and $\mu_i$ for the probabilities that 
$D_b$ changes from $i$ to $i+1$ or $i-1$ in one time step (generation). 
The former outcome occurs if a new allele supplied by mutation gets fixed 
in the population. The latter outcome occurs if an ancestral allele brought 
by immigrants replaces a new allele previously fixed. I disregard the 
possibility of more than one substitution in one time step. Probabilities 
$\lambda_i$ and $\mu_i$ are small and depend on the rate of migration, $m$, 
the rate of mutation per gamete per generation, $v$, the strength of 
selection for local adaptation, $s$, and the population size, $N$. 
Speciation occurs when $d$ hits the (absorbing) boundary $K+1$. If 
this happens, the population is completely reproductively isolated from the 
ancestral genotypes. I do not consider the possibility of backward mutation 
towards an ancestral state. Fixing new alleles at $K+1$ loci completes the 
process of speciation.

\section{3. RESULTS}

I will compute two important characteristics of the speciation process.
The first is the average waiting time to speciation, $t_0$, defined as the
average time to reach the state of complete reproductive isolation ($D_b=K+1$) starting at the ancestral state ($D_b=0$). In general, during the interval 
from $t=0$ to the time of speciation the population will repeatedly 
accumulate a few substitutions only to lose them and return to the ancestral 
state at $D_b=0$. The second
characteristic is the average duration of speciation, $T_0$, defined as the
time that it takes to get from the ancestral state ($D_b=0$) to the state 
of complete reproductive isolation ($D_b=K+1$) {\em without returning to the
ancestral state}. [The duration of speciation is similar to the conditional 
time that a new allele destined to be fixed segregates before fixation.]

\subsection{(a) ALLOPATRIC SPECIATION}

It is illuminating to start with the case of no immigration (cf. Orr 1985;
Orr \& Orr 1996; Gavrilets 1999, pp. 6-8). In this case, the process of 
accumulation of new mutations is irreversible and the average duration 
of speciation, $T_0$, is equal to the average waiting time to speciation, 
$t_0$. \\

{\em No selection for local adaptation}.
With no or very little within-population genetic variation the process
of accumulation of substitutions leading to reproductive isolation is 
effectively neutral (cf., Orr 1995; Orr \& Orr 1996). The average 
number of neutral mutations fixed per generation equals the mutation rate 
$v$ (Kimura 1983). Thus, the average time to fix $K+1$ mutations is 
	\begin{equation}
		t_0 = \frac{K+1}{v}.
	\end{equation}

{\em Selection for local adaptation}.  
In a diploid population of size $N$, the number of mutations per generation is 
$2Nv$. The probability of a mutant allele with a small selective advantage
$s$ being fixed is approximately $2s/(1-exp(-4Ns))$ (Kimura 1983). Thus, the 
average time to fix $K+1$ mutations is 
	\begin{equation}
		t_0 = \frac{K+1}{v}\ \frac{1-exp(-S)}{S},
	\end{equation}
where $S=4Ns$. With $S$ increasing from 0 to, say, 10, the time to speciation $t_0$ decreases to approximately 1/10 of that in the case of no selection
for local adaptation.

\subsection{(b) PARAPATRIC SPECIATION}

With immigration, the dynamics of $D_b$ are controlled by two opposing 
types of forces. Mutation and selection act to increase $D_b$ 
whereas migration acts to decrease $D_b$.
The appendix presents exact formulae for $t_0$ and $T_0$ in the case of
parapatric speciation. Below I give some simple approximations valid 
if $\frac{m}{v}\ exp(-S)$ is not too small.\\

{\bf Threshold function of reproductive compatibility}.
Here, I assume that the function $w(d)$ specifying the probability that two individuals are not reproductively isolated has a threshold form:
    \begin{equation} \label{thres}
         w(d) = \left\{ \begin{array}{cc}
                           1 & \mbox{for $d\leq K$},\\
                           0 & \mbox{for $d>K$}.
                       \end{array}
                \right.
    \end{equation}
(Gavrilets {\em et al.} 1998, 2000{\em ab}; Gavrilets 1999, cf. Higgs and Derrida 1992).
This function implies that immigrants have absolutely no problems 
mating with the residents unless the genetic distance $D_b$ exceeds $K$.
I start with the worst-case scenario for speciation when not only
immigrants can easily mate with residents but also selection for local 
adaptation is absent (cf., Gavrilets {\em et al.} 1998, 2000{\em a}; Gavrilets 1999).

\begin{table}[tbh]

\begin{center}
\begin{tabular}{|c|c|c|c|} 
\hline

\multicolumn{1}{|c|}{ } & \multicolumn{1}{c}{Allopatric case} &  \multicolumn{2}{|c|}{Parapatric case}\\
\cline{3-4}   

 $K$ & ($t_0=T_0$)  & $t^*_0$ & $T^*_0$ \\
\hline  
$1$ & $\frac{2}{v}$	& $(2+R)\frac{1}{v}$ & $ \frac{2+R}{1+R}\ \frac{1}{v}$\\  [.25in]

$2$ & $\frac{3}{v}$ 	& $(3+3R+2R^2)\frac{1}{v}$ & $ \frac{3+4R+2R^2}{1+R+2R^2}\ \frac{1}{v}$\\ [.25in]

$3$ & $\frac{4}{v}$	& $(4+6R+8R^2+6R^3)\frac{1}{v}$ & $ \frac{4+8R+13R^2+6R^3}{1+R+2R^2+6R^3}\ \frac{1}{v}$\\ [.25in]
\hline
\end{tabular}
\end{center}

\caption{ {\bf Exact expressions for the average waiting time to speciation, 
$t^*_0$, and the average duration of speciation, $T^*_0$, for small $K$ with no 
selection for local adaptation and a threshold function of reproductive 
compatibility. $R=m/v$.}}
\end{table}

\begin{figure}[hbt]
\begin{center}
\scalebox{0.85}{\includegraphics[.5in,2in][8in,5.5in]{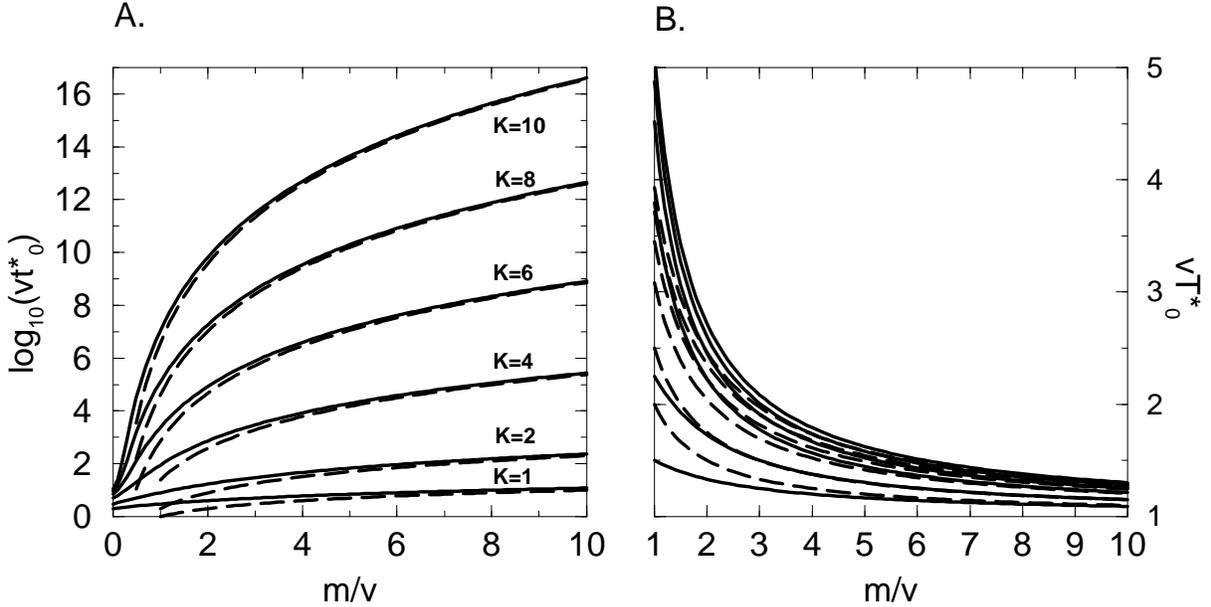}}
\end{center}
\caption{ {\bf The average waiting time to speciation, $t_0^*$, and the 
average duration of speciation, $T_0^*$. The $x$-axes give the ratio of 
migration and mutation rates. In Figure (A), the $y$-axis gives the product 
of $t_0^*$ and the mutation rate $v$ on the logarithmic scale. In Figure (B), 
the $y$-axis gives the product of $T_0^*$ and the mutation rate $v$ on the 
linear scale. Solid lines correspond to the exact values (found in the 
Appendix); dashed lines correspond to the approximate equations 
(\ref{time_t0}) and (\ref{time_T0}). In Figure (B), the lines correspond to 
$K=1,2,4,6,8$, and $10$ (from bottom to top).}}
\end{figure}

{\em No selection for local adaptation}.
With no selection for local adaptation and neglecting within-population 
genetic variation, the process of fixation is approximately neutral. 
The probability of fixation of an allele is equal to its initial frequency. 
The average frequency of new alleles per generation is approximately the
mutation rate $v$. 
If the immigrants differ from the residents at $D_b=i$ loci, there are $i$ 
loci that can fix ancestral alleles brought by migration. The average 
frequency of such alleles per generation is $i m$. Thus, the probabilities 
of stochastic transitions increasing and decreasing $D_b$ by one are 
approximately
	\begin{equation} \label{lambda1}
		\lambda_i = v,\ \mu_i = i\ m.
	\end{equation}
With small $K$, the exact expressions for $t_0$ and $T_0$ found in the
Appendix are relatively 
compact (see Table 1). With larger $K$, the approximate equations are 
more illuminating. The average waiting time to speciation is approximately
	\begin{equation} \label{time_t0}
		t^*_0 \approx  \frac{1}{v} \left( \frac{m}{v} \right)^K K!
	\end{equation}
The average duration of speciation is approximately
	\begin{equation} \label{time_T0}
		T^*_0 \approx  \frac{1}{v} \left(1+ \frac{ \Psi(K+1)+\gamma}{m/v} \right),
	\end{equation}
where $\gamma \approx 0.577$ is Euler's constant and $\Psi(.)$ is the psi
(digamma) function (Gradshteyn \& Ryzhik, 1994). [Function $\Psi(K+1)+\gamma$
slowly increases with $K$ and is equal to 1 at $K=1$, to 2.93 at $K=10$ and
to 5.19 at $K=100$.] For example, if $m=0.01, v=0.001$ and $K=5$,
then the waiting time to speciation is very long: $t^*_0 \approx 1.35 \times 
10^{10}$ generations, but if speciation does happen, its duration is 
relatively short: $T^*_0 \approx 1236$ generations. Figure 1 illustrates 
the dependence of $t^*_0$ and $T^*_0$ on model parameters. Notice that 
$T^*_0$ is order $1/v$ across a wide range of parameter values.\\

\begin{figure}[tbh]
\begin{center}
\scalebox{0.85}{\includegraphics[.5in,2in][8in,5.in]{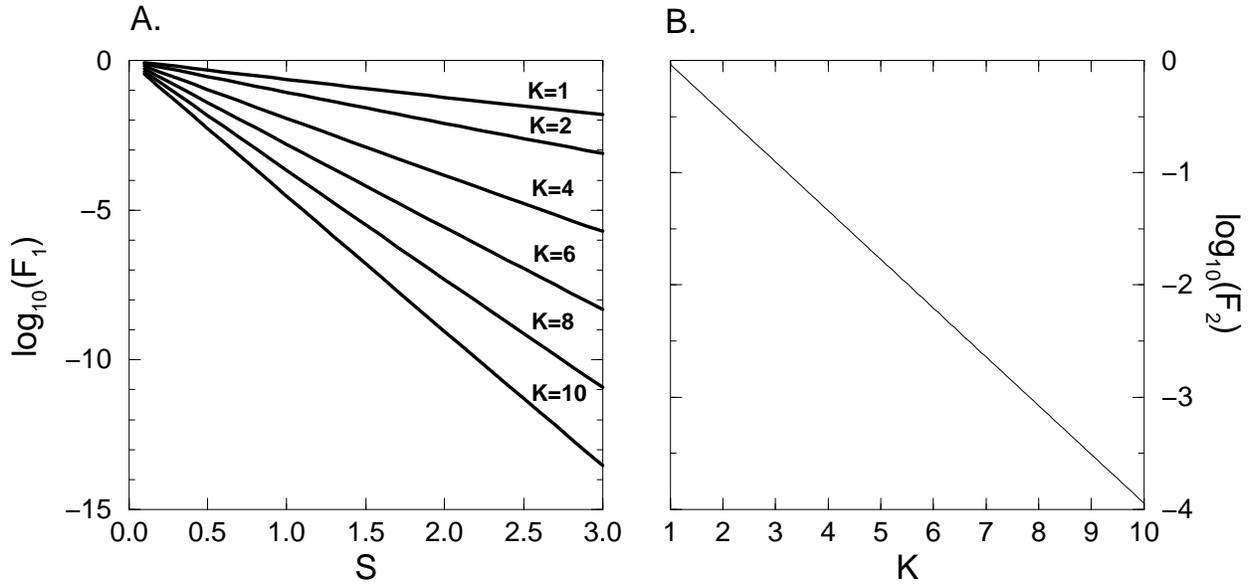}}
\end{center}
\caption{ {\bf Effects of selection for local adaptation and of linear 
function of reproductive compatibility on the 
waiting time to speciation. (A) The proportion 
(on the logarithm scale) by which selection for local adaptation
decreases $t_0$ ($F_1=exp(-KS)\ \frac{1-exp(-S)}{S}$; see equation
\ref{F1}). (B) The proportion
(on the logarithm scale) by which $t_0$ is reduced relative
to $t_0^*$ if the function of reproductive compatibility is linear 
($F_2=\sqrt{2 \pi}\ exp(-K)$; see equation \ref{F3}). 
}}
\end{figure}

{\em Selection for local adaptation}.
Assume that ``new'' alleles improve adaptation to the local conditions. Let 
$s$ be the average selective advantage of a new allele over the corresponding
ancestral 
allele. Each generation there are $2Nv$ such alleles supplied by mutation. 
The probability of fixation of an advantageous allele is approximately 
$2s/[1-exp(-4Ns)]$. Migration brings approximately $2Nmi$ ancestral alleles 
at the loci that have previously fixed new alleles. These alleles are 
deleterious in the new environment. The probability of fixation of a 
deleterious allele is approximately $2s/[exp(4Ns)-1]$ (Kimura 1983). Thus, 
the probabilities of stochastic transitions increasing and decreasing $D_b$ 
by one are approximately
	\begin{equation} \label{lambda2}
		\lambda_i = v\  \frac{4Ns}{1-exp(-4Ns)},\ 
		\mu_i = i\ m\  \frac{4Ns}{exp(4Ns)-1}.
	\end{equation}
The waiting time to speciation is approximately 
	\begin{equation} \label{F1}
		t_0 \approx t_0^*\ exp(-KS)\ \frac{1-exp(-S)}{S},
	\end{equation}
where $t_0^*$ is given by equation (\ref{time_t0}). The average duration 
of speciation is approximately
	\begin{equation} \label{F2}
		T_0 \approx  \frac{1}{v} \left(1+ \frac{ \Psi(K+1)+\gamma}{(m/v)e^{-S}} \right)\ \frac{1- exp(-S)}{S}.
	\end{equation}
For example, if $m=0.01, v=0.001, K=5$ and $S=2$, then $t_0 \approx 2.74 
\times 10^4$ generations and $T_0=2170$ generations. Thus, selection for 
local adaptation dramatically decreases $t_0$ (in the numerical 
example, by the factor $\approx$ 50,000) and somewhat increases $T_0$ 
relative to the case of speciation driven by mutation and genetic drift.
Figure 2A illustrates the effect of selection for local 
adaptation on $t_0$ in more detail.\\

{\bf Linear function of reproductive compatibility}.
Here, I assume that the probability of no reproductive isolation decreases
linearly with genetic distance $d$ from one at $d=0$ to zero at $d=K+1$:
    \begin{equation} \label{linear}
         w = \left\{ \begin{array}{cc}
                           1-i/(K+1) & \mbox{for $d\leq K$},\\
                           0 & \mbox{for $d>K$}.
                       \end{array}
                \right.
    \end{equation}
Now, immigrants experience problems in finding compatible mates even
when the genetic distance is below $K+1$.\\

{\em No selection for local adaptation}.
With no selection for local adaptation, the probabilities of stochastic 
transitions $\lambda_i $ and $\mu_i$ are given by equations (\ref{lambda1})
with $m$ substituted for an ``effective'' migration rate
	\begin{equation} \label{m_e}
		m_i=m \left( 1- \frac{i}{K+1} \right).
	\end{equation}
The waiting time to speciation is approximately  
	\begin{equation} \label{F3}
		t_0 \approx t_0^*\ \sqrt{2 \pi}\ exp(-K),
	\end{equation}
where $t_0^*$ is given by equation (\ref{time_t0}). 
The average duration of speciation is approximately 
	\begin{equation} \label{T**}
		T_0 \approx \frac{1}{v} \left(1+2\
\frac{ \Psi(K+1)+\gamma }{m/v} \right).
	\end{equation}
where $\gamma$ is Euler's constant and $\Psi(.)$ is the psi (digamma) function.
The last equation differs from equation (\ref{time_T0}) only by the factor 2
inside the parentheses. 
For example, with the same values of parameters as above $t_0 \approx 2.36
\times 10^8$ generations and $T_0 = 1470$ generations. Thus, $t_0$ is 
significantly reduced (by the 
factor 57) whereas $T_0$ is somewhat larger than in the case of threshold 
function of reproductive compatibility.  
Figure 2B illustrates the effect of linear function of reproductive 
compatibility on $t_0$ in more detail.\\

{\em Selection for local adaptation}.
With selection for local adaptation, the probabilities of stochastic 
transitions $\lambda_i $ and $\mu_i$ are given by equations (\ref{lambda2})
with $m$ substituted for an ``effective'' migration rate (\ref{m_e}).
The average time to speciation is approximately
	\begin{equation}
		t_0 \approx t_0^*\ \sqrt{2 \pi}\ exp(-K)\ exp(-KS)\ \frac{1-exp(-S)}{S}.
	\end{equation}
The average duration of speciation $T_0$ is given by equation (\ref{F2}) 
with an additional factor 2 placed in front of the ratio in the parentheses.
As before, selection for local adaptation substantially decreases $t_0$ and
slightly increases $T_0$.

\section{4. DISCUSSION}

The results presented above allow one to get insights about the
time scale of parapatric speciation driven by mutation, random genetic 
drift and/or selection for local adaptation. I start the discussion of
these results by considering the original Dobzhansky model. 

\subsection{(a) Two-locus two-allele Dobzhansky model}

Dobzhansky's original model (Dobzhansky 1937) describes a two-locus two-allele
system where a specific pair of alleles is incompatible in the sense that the interaction of these alleles ``produces one of the physiological isolating
mechanisms'' (p. 282). Let us assume that the immigrants have ancestral
haplotype {\bf ab} and that the derived allele {\bf B} is incompatible with
the ancestral allele {\bf a} (see Figure 4). In this case, the population 
can evolve to a
state reproductively isolated from the ancestral state via a state with
haplotype {\bf Ab} fixed: ${\bf ab \rightarrow Ab \rightarrow AB}$. 
[Recently Johnson {\em et al.} (2000) considered the probability of parapatric speciation driven by mutation in a somewhat similar model. However, their 
major equation (eq. 13) is heuristic and does not appear to be justified.] 
Let $\nu$ be the probability of mutation from an ancestral allele ({\bf a} 
and {\bf b}) to the corresponding derived allele ({\bf A} or {\bf B}). The 
average waiting time to and the average duration of parapatric speciation 
in this system are given by our general equations with $K=1$ and $v=\nu$. 
Allowing for equal selective advantage $s$ of derived alleles over 
the ancestral alleles,
	\begin{subequations}
	\begin{eqalignno}
		t_0 & =  {\frac {\left (2\,\nu+m{e^{-S}}\right )\left (1-{e^{-S}}\right )}{{\nu}^{2}S}} \approx  \frac{m}{\nu^2}\
\frac{1-e^{-S}}{S}\ e^{-S},\\
		T_0 & =  \frac{ (2 \nu+m e^{-S})(1-e^{-S})}
{\nu S (\nu+me^{-S})} \approx \frac{1}{\nu}\
 \frac{1-e^{-S}}{S},
	\end{eqalignno}
	\end{subequations}
where $S=4Ns$ and the approximations are good if $(m/\nu)exp(-S)>>1$. 
With no selection for local adaptation (that is if $S=0$), 
$t_0 \approx m/\nu^2, T_0 \approx 1/\nu$.

\begin{figure}[tbh]
\begin{center}
\scalebox{0.5}{\includegraphics[.5in,5in][8in,9.5in]{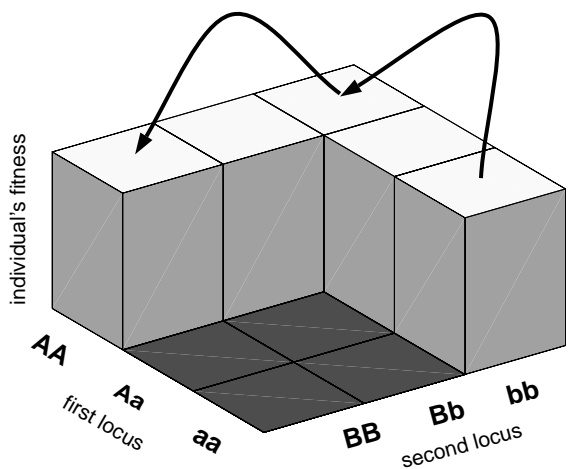}}
\end{center}
\caption{ {\bf Adaptive landscape in the two-locus two-allele Dobzhansky
model. Alleles {\bf a} and {\bf B} are incompatible.
The arrows specify the chain of gene substitutions leading to 
complete reproductive isolation.}}
\end{figure}

Let $m=0.01$ and $\nu=10^{-5}$. Then with no selection for local adaptation, 
the average waiting time to speciation is very long: $t_0 
\approx 10^8$ generations, and $T_0 \approx 10^5$ generations. 
However, even with relatively weak selection for local 
adaptation, $t_0$ can decrease by 1-2 orders of magnitude. For example, 
with $S=1$, $t_0 \approx 2.34 \times 10^7,\ T_0 \approx 6.34 \times 10^4$; 
with $S=2$, $t_0 \approx 5.94 \times 10^6,\ T_0 \approx 4.36 \times 10^4$; 
and with $S=3$, $t_0 \approx 1.64 \times 10^6,\ T_0 \approx 3.23 \times 10^4$.
Because the waiting time to speciation in the two-locus Dobzhansky model 
scales as one over the mutation rate per locus squared, this time is rather 
long. However, the overall number of loci involved in the initial stages of
reproductive isolation is at least on the order of tens to hundreds 
(e.g. Singh 1990; Wu \& Palopoli 1994; Coyne \& Orr 1998; Naveira and 
Masida 1998). This increases the overall mutation rate and can make 
speciation much more rapid. \\

\subsection{(b) Average waiting time to parapatric speciation}

In the models studied here, reproductive isolation is a consequence of cumulative genetic divergence over a set of loci potentially affecting 
mating behavior, fertilization processes, and/or offspring viability and fertility. The underlying biological intuition is that organisms that are 
reproductively compatible should not be too different genetically.
Most species consist of geographically structured populations, some of which
experience little genetic contact for long periods of time (Avise 2000).
Different mutations are expected to appear first and increase in frequency
in different populations  
necessarily resulting in some geographic differentiation even without any
variation in local selection regimes. An interesting question is whether
mutation and drift alone are sufficient to result in parapatric speciation.
This question is particularly important given a
growing amount of data suggesting that rapid evolution of reproductive 
isolation is possible without selection for local adaptation involved
(e.g. Vacquier 1998; Palumbi 1998; Howard 1999).
Our results provide an affirmative answer to this question (see also
Gavrilets {\em et al.} 1998, 2000{\em a}; Gavrilets 1999). However, here the waiting time
to speciation is relatively short only if a very small number of genetic 
changes is sufficient for complete reproductive isolation. For example, 
$t_0$ is on the order of 10-1000 times the inverse of the mutation rate if 
$K=1$ or $2$ with a threshold function of reproductive compatibility, and if 
$K=1, 2$ or $3$ with a linear function of reproductive compatibility. 
It is well recognized that selection for local adaptation can result in
speciation in the presence of some gene flow (e.g. Slatkin 1981; Rice 1984; 
Rice \& Salt 1988; Schluter 1998). Our results show that even relatively
weak selection can dramatically reduce the waiting time to speciation
by orders of magnitude (see Figure 2a). 

\subsection{(c) How much migration prevents speciation?}

In general, evolutionary biologists accept that very small levels of migration
are sufficient to prevent any significant genetic differentiation of 
the populations not to mention speciation (e.g. Slatkin 1987, but see
Wade \& McCauley 1984). To a large degree, this belief appears to be based 
on two observations. One is
that the expected value of the fixation index $F_{ST}$ is small even with a single migrant per generation (e.g. Hartl \& Clark 1997). Another is that 
the expected distribution of allele frequency in the island model changes 
from a U-shaped (which implies at least some genetic differentiation) to a 
bell-shaped (which implies no genetic differentiation on average) as the 
average number of migrants become larger than one per generation
(e.g. Crow \& Kimura 1970). 
However, the equilibrium expectations derived under neutrality theory can be rather misleading if there is
a possibility for evolving complete reproductive isolation. For example, in 
the model with no selection for local adaptation considered above the expected 
change per generation in the genetic distance $D_b$ between the immigrants and 
residents is
	\[
		\Delta D_b =v-mD_b,
	\]
where the first term describes an expected increase in $D_b$ because of new 
mutations and the second term describes an expected decrease in $D_b$ because 
of the influx of ancestral genotypes. This equation predicts that $D_b$ 
will reach an equilibrium value of $v/m$. From this one can be tempted to 
conclude that unless the migration rate is smaller than that of mutation 
($v>m$), $D_b$ cannot be larger than one and, thus, no speciation is possible.
However, this argument is flawed.  Because of the inherent stochasticity 
of the system there is always a non-zero probability of $D_b$ moving any pre-specified distance from 0 which will lead to reproductive isolation.

Strictly speaking, in the models studied here migration does not {\em prevent}
but rather {\em delays} speciation. [The resulting delay can be substantial 
and for all practical reasons infinite.]
For definiteness, I will say that speciation is {\em effectively prevented} 
if the average waiting time to speciation is larger than 1000 times the 
inverse of the mutation rate (that is if $\log_{10}(vt_0)>3$). If the number of 
genetic substitutions necessary for speciation is small (for example, $K=1$, 
as in the original Dobzhansky model, or $K=2$), then migration rates higher 
than 
$10v$ will effectively prevent speciation in the absence of selection for 
local adaptation. For 
example, if $v=10^{-3}$, then speciation is possible with $m$ as high as 0.01.
However, if $v=10^{-5}$, then any migration rate higher than 0.0001 will 
effectively prevent speciation.
If the number of genetic changes required for speciation is relatively large,
say, if $K=10$, then without selection for local adaptation speciation is
effectively prevented (see Fig. 4a). However, relatively weak selection, 
say with $S=2.5$ would overcome migration rates as high as $10v$ if the 
strength of 
reproductive isolation increases linearly with genetic distance (see Fig.4b).

Within the modeling framework used, all immigrants had a fixed genetic composition
which did not change in time. Alternatively, one can imagine two populations exchanging migrants assuming that both populations can evolve. If there is 
no selection for local adaptation, this case is mathematically equivalent 
to that studied above but with the mutation rate being twice as large as in 
the case of a single evolving population. Therefore, the waiting time to
speciation in the two population case will dramatically decrease relative
to that in the single population case. The maximum migration rates 
compatible with speciation will be twice as large as before.

\begin{figure}[tbh]
\begin{center}
\scalebox{0.85}{\includegraphics[.5in,2in][8in,5.5in]{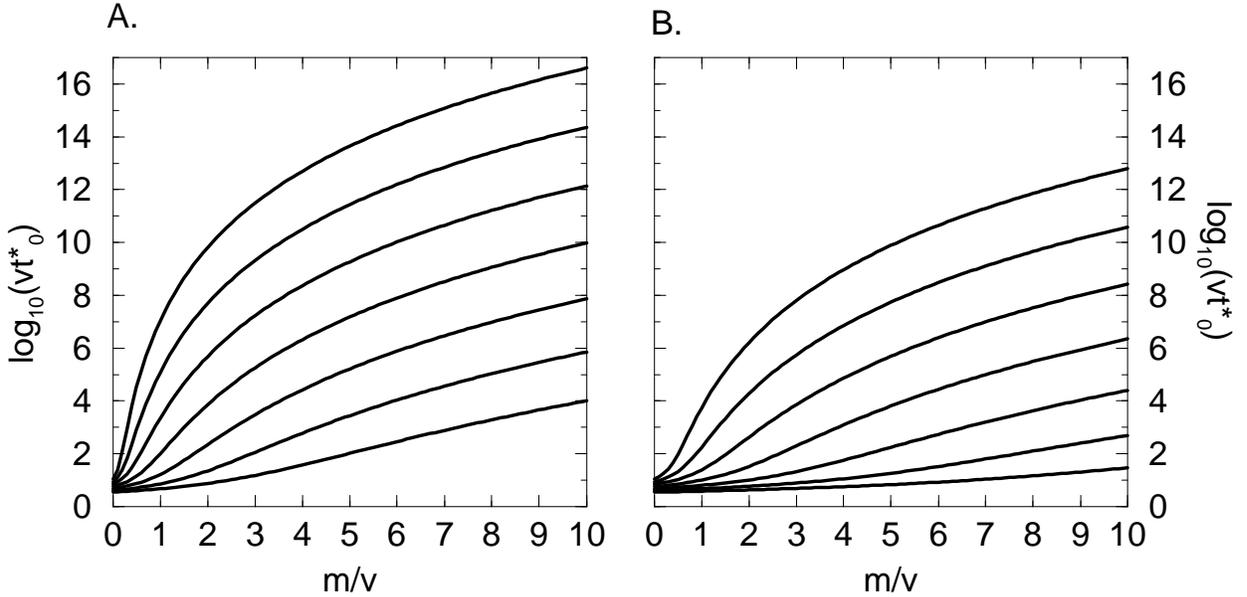}}
\end{center}
\caption{ {\bf Waiting time to speciation with selection for local adaptation
for $K=10$. Different lines correspond to $S=0.0, 0.5, 1.0, 1.5, 2.0, 2.5$, 
and $3.0$ (from top to bottom). (A) Threshold function of reproductive
compatibility (equation \ref{thres}). (B) Linear function of reproductive
compatibility (equation \ref{linear}).}}
\end{figure}

\subsection{(c) The role of environment}

The waiting time to speciation, $t_0$, is extremely sensitive to parameters:
changing a parameter by a small factor, say two or three, can increase or
decrease $t_0$ by several orders of magnitude. Looking across a range of
parameter values, $t_0$ is either relatively short (if the parameters are 
right) or effectively infinite. Most of the parameters of the model (such
as the migration rate, intensity of selection for local adaptation, the
population 
size, and, probably, the mutation rate) directly depend on the state of the 
environment (biotic and abiotic) the population experiences. This suggests
that speciation can be triggered by changes in the environment (cf.
Eldredge 2000). Note that the time lag between an environmental change initiating speciation and an actual attainment of reproductive isolation 
can be quite substantial as our model shows. If it is an 
environmental change 
that initiates speciation, the populations of different species inhabiting 
the same geographic area should all be affected. In this case, 
one expects more or less synchronized bursts of speciation in a geographic
area - that is a ``turnover pulse'' (Vrba 1985).

\subsection{(d) Average duration of parapatric speciation}

In our model, the average waiting time to and the average duration of 
allopatric speciation are identical. Previously, Lande (1985) and Newman 
{\em et al.} (1985) studied how an isolated population can move from one adaptive
peak to another by random genetic drift. They showed that the average duration 
of stochastic transitions between the peaks is much shorter than the time 
that the population spends in a neighborhood of the initial peak before the 
transition. Within the framework used by these authors stochastic transitions 
are possible in a reasonable time only if the adaptive valley separating the
peaks is shallow. This implies that reproductive isolation resulting from a single transition is very small. Potentially, strong or even complete
reproductive isolation (that is speciation)
can result from a series of peak shifts along a chain 
of ``intermediate'' adaptive peaks such that each individual transition is across a shallow valley but the cumulative effect of many peak shifts is
large (Walsh 1982). In this case, the results of Lande (1985) and Newman 
{\em et al.} (1985) actually imply that the population will spent a very long 
time at each of
the intermediate adaptive peaks. This would lead to a very long duration of
allopatric speciation that is in fact comparable to the overall waiting 
time to speciation. 

For parapatric speciation, the predictions are very different.
Our results about the duration of speciation lead to 
three important generalizations. The first 
is that the average duration of parapatric speciation, $T_0$, is much smaller 
than the average waiting time to speciation, $t_0$. This feature of the 
models studied here is compatible with the patterns observed in the fossil 
record which form the empirical basis of the theory of punctuated equilibrium 
(Eldredge 1971; Eldredge \& Gould 1972).  
The second generalization concerns the absolute value of $T_0$.
The waiting time to speciation changes dramatically with slight changes in 
parameter values. In contrast, the duration of speciation is on the order
of one over the mutation rate over a subset of the loci affecting reproductive 
isolation for a wide range of migration rates, population 
sizes, intensities of selection for local adaptation, and the number of
genetic changes required for reproductive isolation. Given a ``typical'' 
mutation rate on the order of $10^{-5}-10^{-6}$ per locus per generation
(e.g. Griffiths {\em et al.} 1996; Futuyma 1997) and assuming that there are at 
least on the order of 10-100 genes involved in the initial stages of the
evolution of reproductive isolation (e.g. Singh 1990; Wu \& Palopoli 1994; 
Coyne \& Orr 1998; Naveira \& Masida 1998), the duration of
speciation is predicted to range between $10^3$ and $10^5$ generations with
the average on the order of $10^4$ generations. The third generalization
is about the likelihood of situations where strong but not complete 
reproductive isolation between populations is maintained for an extended 
period of time (much longer than the inverse of the mutation rate) in the presence of 
small migration without the populations becoming completely isolated or 
completely compatible. Judging from our theoretical results, such situations appear to be extremely improbable.

\subsection{(e) Validity of the approximations used}

The results presented here are based on a number of approximations the most
important of which is the assumption that within-population genetic variation
in the loci underlying reproductive isolation can be neglected. 
A biological scenario to which this assumption is most 
applicable is that of a small (peripheral) populations with not much genetic
variation maintained and with occasional influx of immigrants from the main 
population. 
[ Note that within-population genetic 
variation in the loci underlying reproductive isolation has to be manifested 
in reproductive incompatibilities between some members of the population.
However, the overall proportion of incompatible mating pairs within the 
population is not expected to be large (e.g. Wills 1977; Nei {\em et al.}1983;
Gavrilets 1999).]
Intuitively, one might expect that increasing within-population
variation would substantially increase the rate of substitutions by random genetic drift
and make speciation easier. However, in polymorphic populations the alleles 
affecting the degree of reproductive isolation cannot be treated as neutral 
because they are weakly selected against than rare (Gavrilets {\em et al.} 
1998, 2000{\em a}; Gavrilets 1999). In the absence of selection for local 
adaptation this might make speciation somewhat more difficult. 
Allowing for genetic variation among immigrants can increase the plausibility
of speciation. For example, if new alleles are deleterious in the ancestral environment and are maintained there by mutation, their equilibrium frequency will be order $v/s^*$, where $s^*$ is the selection coefficient against
new alleles in the ancestral environment. Thus, the overall frequency of 
new alleles in the population per generation will increase from $v$ to
approximately $v+m \frac{v}{s^*}$. Intuitively, this can result in a 
substantial reduction in the waiting time to speciation.
The overall effect of genetic variation (both within-population and among immigrants) on the waiting time to parapatric speciation has to be 
explored in a systematic way. This is especially important given that the 
individual-based simulations reported in Gavrilets {\em et al.} (1998, 2000a) 
show that rapid speciation is possible well beyond the domain of parameter values identified here as conducive to speciation.
As for the duration of speciation, I expect it to have an order of one over the 
level of genetic variation maintained in the loci underlying reproductive 
isolation. As such, with genetic variation the duration of speciation is
expected to be (much) shorter than $1/v$.\\

I am grateful to Janko Gravner for helpful suggestions and to
Chris Boake for comments on the manuscript.
This work is supported by National Institutes of Health grant 
GM56693 and by EPPE fund (University of Tennesee, Knoxville).\\

\section{APPENDIX}

{\bf Average waiting time till speciation}.
I consider a Markov chain with $K+1$ states $0,1,\dots,K,K+1$. Let $p_{ij}$
be the corresponding transition probabilities. I assume 
that the state $K+1$ is absorbing but the state 0 is not. Let $t_i$ be the 
average time till absorption starting from $i$. The mean absorption times 
satisfy to a general system of linear equations
	\begin{equation} \label{gen}
		t_i=1+\sum_j p_{ij}t_j
	\end{equation}
for $i=0,1,\dots,K$ with $t_{K+1}=0$ (e.g. Norris 1997). I assume that the transition probabilities are $p_{i,i+1}=\lambda_i, p_{i,i-1}=\mu_i$ and 
$p_{ij}=0$ if $|i-j|>1$ with $\mu_{K+1}=0$. In this case, the system of linear equations 
(\ref{gen}) can be solved by standard methods (e.g. Karlin \& Taylor 1975). 

Let $z_i=t_i-t_{i+1}$. From equation (\ref{gen}) with $i=0$ one finds an 
equality $t_0=1+\lambda_0 t_1+(1-\lambda_0)t_0$ which can be rewritten as
	\begin{subequations} \label{rec}
	\begin{equation}
		z_0=1/\lambda_0.\label{rec-a}
	\end{equation}
In a similar way, for $i>0$ one finds an equality $t_i=1+\lambda_i t_{i+1}
+\mu_i t_{i-1}+(1-\lambda_i-\mu_i)t_i$ which can be rewritten as
	\begin{equation}
		z_i=\frac{\mu_i}{\lambda_i}z_{i-1}+\frac{1}{\lambda_i}. \label{rec-b}
	\end{equation}
	\end{subequations}
The solution of the system of linear recurrence equations (\ref{rec}) is
	\begin{subequations} \label{sol}
	\begin{equation}
		z_i = \frac{\rho_i}{\lambda_0}+\sum_{j=1}^{i} 
\frac{\rho_i}{\lambda_j \rho_j}, \label{sol-a}
	\end{equation}
where
	\begin{equation}
		\rho_j= \frac{\mu_1 \mu_2 \dots \mu_j}{\lambda_1 \lambda_2 
\dots \lambda_j} \label{sol-b}
	\end{equation}
	\end{subequations}
with $\rho_0=1$. One can also see that $\sum_{i=0}^K z_i=(t_0-t_1)+(t_1-t_2)+
\dots (t_K-t_{K+1})=t_0$. Thus,  $t_0$ can be found by summing up equations
(\ref{sol-a}):
	\begin{equation} \label{t_0}
		t_0 = \frac{ \sum_{i=0}^K \rho_i}{\lambda_0}+ 
\sum_{i=1}^K \sum_{j=1}^{i} \frac{\rho_i}{\lambda_j \rho_j}.
	\end{equation}
The absorption times $t_i$ corresponding to $i>0$ can be found recursively
using equation (\ref{rec-b}).

With a threshold function of reproductive compatibility (\ref{thres}),
	\begin{equation}
		\rho_j = R^j j!,
	\end{equation}
where $R=(m/v)exp(-S)$.
With a linear function of reproductive compatibility (\ref{linear}),
	\begin{equation}
		\rho_j = R^j j!\ \frac{K!}{(K+1)^j (K-j)!}.
	\end{equation}

{\bf Average duration of speciation}.
The average duration of speciation, $T_0$, can be defined as the average time 
that it takes to walk from state $0$ to state $K+1$ without returning to 
state $0$. Ewens (1979, Section 2.11) provides formulae that can be used 
to find $T_0$. These formulae are summarized below.

The probability of entering state $K+1$ before state $0$ starting from $i$ is
	\begin{equation}
		\pi_i = \sum_{j=0}^{i-1} \rho_j /\ \sum_{j=0}^{K} \rho_j.
	\end{equation}
Starting from state $i$, the mean time spent in state $j$ before entering 
state $0$ or state $K+1$ is
	\begin{subequations}
	\begin{eqalignno}
		t_{ij} & =(1-\pi_i) \sum_{k=0}^{j-1}\rho_k/(\rho_j \lambda_j)\ {\rm
for}\ j=1,\dots, i,\\
		t_{ij} & = \pi_i \sum_{k=j}^{K}\rho_k/ (\rho_j \lambda_j)
\hspace{.5in} {\rm for}\ j=i+1,\dots, K.
	\end{eqalignno}
	\end{subequations}
Starting from state $i$, the conditional mean time spent in state $j$
for those cases for which the state $K+1$ is entered before state $0$ is
	\begin{equation}
		t^*_{ij} = t_{ij} \pi_j/\pi_i.
	\end{equation}
The condition mean time till absorption in $K+1$ is 
	\begin{equation}
		t^*_i=\sum_{j=1}^K t^*_{ij}. 
	\end{equation}

The average duration of speciation is the sum of the average
time spent in state $0$ before moving to state $1$, which is $1/\lambda_0$,
plus the conditional mean time till absorption in $K+1$ starting from state
$1$, which is $t^*_1$,
	\begin{equation}
		T_0= 1/\lambda_0+ t^*_1.
	\end{equation}

\section{REFERENCES}

\begin{enumerate}

\item[]\label{avi}Avise, J. C. 2000 {\em Phylogeography}. Harvard 
University Press, Cambridge, Massachusetts.

\item[]\label{bar93} Barton, 1993 The probability of fixation of a favoured allele 
in a subdivided population. {\em Genet. Res.} {\bf 62}, 149-157.

\item[]\label{bur95} Burger, W. 1995 Montane species-limits in Costa Rica
and evidence for local speciation on attitudinal gradients. In
{\em Biodiversity and conservation of neotropical montane forests} (S. P.
Churchill, ed.), pp. 127-133. New York:The New York Botanical Garden.

\item[]\label{coy98} Coyne, J. A. \& H. A. Orr. 1998 The evolutionary genetics 
of speciation. {\em Phil. Trans. R. Soc. Lond. B} {\bf 353}, 287-305.

\item[]\label{cro70} Crow, J. F. \& Kimura, M. 1970 {\em An introduction
to population genetics theory}. Minneapolis, Minnesota: Burgess Publishing Company.

\item[]\label{die99} Dieckmann, U. \& Doebeli, M. 1999 On the origin of 
species by sympatric speciation. {\em Nature}, {\bf 400}, 354-357.

\item[]\label{dob37} Dobzhansky, Th. G. 1937 {\em Genetics and the origin of 
species}. New York: Columbia University Press. 

\item[]\label{eld71} Eldredge, N. 1971 The allopatric model and phylogeny in 
Paleozoic invertebrates. {\em Evolution} {\bf 25}, 156-167.

\item[]\label{eld00} Eldredge, N. 2000 The sloshing bucket: how the physical 
realm controls evolution. Pp. 000-000 in Crutchfield, J. and P. Schuster 
(eds.)  
{\em Towards a Comprehensive Dynamics of Evolution - Exploring
the Interplay of Selection, Neutrality, Accident, and Function}.
Oxford University Press. (In the press.)

\item[]\label{eld72} Eldredge, N. \& Gould, S. J. 1972 Punctuated equilibria: an 
alternative to phyletic gradualism. In {\em Models in paleobiology} 
(ed. Schopf, T. J.), pp. 82-115. San Francisco: Freeman, Cooper.

\item[]\label{end77} Endler, J. A. 1977 {\em Geographic variation, speciation 
and clines}.  Princeton, NJ: Princeton University Press. 

\item[]\label{ewe79} Ewens, W. J. 1979 {\em Mathematical population genetics}. 
Berlin: Springer-Verlag.
 
\item[]\label{fri98} Frias, D. \& Atria, J. 1998 Chromosomal variation,
macroevolution and possible parapatric speciation in {\em Mepraia spinolai}
(Porter) (Hemiptera: Reduviidae). {\em Genetics and Molecular Evolution}
{\bf 21}, 179-184.

\item[]\label{fri97} Friesen, V. L. \& Anderson, D. J. 1997. Phylogeny and
evolution of the Sulidae (Aves: Pelecaniformes): a test of alternative modes
of speciation. {\em Molecular Phylogenetics and Evolution}, {\bf 7}, 252-260. 

\item[] \label{Fut97} Futuyma, D. J. 1997 {\em Evolutionary biology}.
Sunderland, MA: Sinauer.

\item[]\label{gav97} Gavrilets, S. 1997 Evolution and speciation on holey 
adaptive landscapes.  {\em Trends Ecol. Evol.} {\bf 12}, 307-312.

\item[]\label{gav99} Gavrilets, S. 1999 A dynamical theory of speciation on 
holey adaptive landscapes. {\em Amer. Natur.} {\bf 154}, 1-22.

\item[]\label{gav00} Gavrilets, S. 2000. "Evolution and speciation in a 
hyperspace: the roles of neutrality, selection, mutation and random drift."  
In Crutchfield, J. and P. Schuster (eds.)  
{\em Towards a Comprehensive Dynamics of Evolution - Exploring
the Interplay of Selection, Neutrality, Accident, and Function}.
Oxford University Press. (In the press.)

\item[]\label{gav98}  Gavrilets, S. \& Boake, C. R. B. 1998 On the evolution 
of premating isolation after a founder event. {\em Amer. Natur.} {\bf 152},
706-716.

\item[]\label{gavgra97} Gavrilets, S. \& Gravner, J. 1997 Percolation on the fitness 
hypercube and the evolution of reproductive isolation. {\em J. Theor. Biol.} 
{\bf 184}, 51-64.

\item[]\label{gav96} Gavrilets, S. \& Hastings, A. 1996 Founder effect speciation: a theoretical reassessment. {\em Amer. Nat.} {\bf 147}, 466-491.

\item[]\label{gav98} Gavrilets, S., Li, H. \& Vose, M. D. 1998 Rapid parapatric 
speciation on holey adaptive landscapes.  {\em Proc. Roy. Soc. Lond.} B 
{\bf 265}, 1483-1489.

\item[]\label{gav00a} Gavrilets, S., Li, H. \& Vose, M. D. 2000{\bf a}
Patterns of parapatric speciation. {\em Evolution} {\bf 54}, 000-000. 

\item[]\label{gav00b} Gavrilets, S., Acton, R. \& Gravner, J. 2000{\bf b}
Dynamics of speciation and diversification in a metapopulation. {\em Evolution} 
{\bf 54}, 000-000.

\item[]\label{gra94} Gradshteyn, I. S. \& Ryzhik, I. M. 1994 {\em Tables of
Integrals, Series, and Products}. San Diego: Academic Press.

\item[]\label{gri96} Griffiths, A. J. F., Miller, J. H., Suzuki, D. T.,
Lewontin, R. C. \& Gelbart,W. M. 1996
{\em An introduction to genetic analysis}. 6th ed.
New York: W. H. Freeman.

\item[]\label{har97} Hartl, D. L. \& Clark, A. G. 1997 {\em Principles of
population genetics}. Sunderland, Massachusetts: Sinauer Associates, .

\item[]\label{hig92}  Higgs, P. G. \& Derrida, B. 1992
Genetic distance and species
formation in evolving populations. {\em J. Mol. Evol.} {\bf 35}, 454-465.

\item[]\label{how99} Howard, D. J. 1999 Conspecific sperm and pollen precedence and 
speciation. {\em Annu. Rev. Ecol. Syst.} {\bf 30}, 109-132.

\item[]\label{joh00} Johnson, K. P., Adler, F. R. \& Cherry, J. L. 2000.
Genetic and phylogenetic consequences of island biogeography. {\em Evolution}
{\bf 54}, 387-396.

\item[]\label{kar75} Karlin, S. \& Taylor, H. M. 1975 {\em A first course in stochastic processes}, second ed.  San Diego: Academic Press.

\item[]\label{kim83} Kimura, M. 1983
{\em The neutral theory of molecular evolution }.  
New York : Cambridge University Press.

\item[]\label{lan79} Lande, R. 1979  Effective deme size during long-term evolution 
estimated from rates of chromosomal rearrangement. {\em Evolution} {\bf 33},
234-251.

\item[]\label{lan85a} Lande, R. 1985 The fixation of chromosomal rearrangements in a subdivided
population with local extinction and colonization. {\em Heredity} {\bf 54},
323-332.

\item[]\label{lan85b} Lande, R. 1985 Expected time for random genetic drift of a population
between stable phenotypic states. {\em Proc. Natl. Acad. Sci. USA} {\bf 82},
7641-7645.

\item[]\label{mac98} Macnair, M. R. \& Gardner, M. 1998 The evolution of
edaphic endemics. Pp. 157-171 in {\em Endless Forms: Species and Speciation} 
(eds Howard, D. J. \& Berlocher, S. H.). 
New York: Oxford University Press.

\item[]\label{nav98} Naveira, H. F. \& Masida, X. R. 1998 The genetic of hybrid male 
sterility in {\em Drosophila}. In {\em Endless Forms: Species and Speciation} 
(eds Howard, D. J. \& Berlocher, S. H.), pp. 330-338. 
New York: Oxford University Press.

\item[] \label{Nei83} Nei, M., Maruyama, T. \& Wu, C.-I. 1983 Models of 
evolution of reproductive isolation.  {\em Genetics} {\bf 103}, 557-579.

\item[]\label{new85} Newman, C. M., Cohen, J. E. \& Kipnis, C. 1985 
Neo-Darwinian 
evolution implies punctuated equilibria. {\em Nature} {\bf 315}, 400-401.

\item[]\label{nor97} Norris, J. R. 1997 {\em Markov Chains}.  Cambridge: 
Cambridge University Press.

\item[]\label{orr95} Orr, H. A. 1995  The population genetics of 
speciation: the evolution of hybrid incompatibilities. {\em Genetics}
{\bf 139}, 1803-1813.

\item[]\label{orr96} Orr, H. A. \& Orr, L. H. 1996 Waiting for speciation: the effect
of population subdivision on the waiting time to speciation. {\em Evolution}
{\bf 50}, 1742-1749.

\item[]\label{pal98} Palumbi, S. R. 1998 Species formation and the evolution of gamete 
recognition loci. In {\em Endless forms: species and 
speciation.} (eds Howard, D.J. \& Berlocher, S. H.), pp. 271-278. New York:
Oxford University Press.

\item[]\label{ric84} Rice, W. R. 1984 Disruptive selection on habitat preferences and
the evolution of reproductive isolation: a simulation study. {\em Evolution}
{\bf 38}, 1251-1260.
 
\item[]\label{ric93} Rice, W. R. \& Hostert, E. E. 1993 Laboratory experiments on
speciation: what have we learned in 40 years? {\em Evolution} {\bf 47},
1637-1653.

\item[]\label{ric88} Rice, W. R. \& Salt, G. B. 1988 Speciation via disruptive selection
on habitat preference: experimental evidence. {\em Amer. Natur.} {\bf 131},
911-917.

\item[]\label{rip90} Ripley, S. D. \& Beehler, B. M. 1990. Patterns of
speciation in Indian birds. {\em J. of Biogeography} {\bf 17}, 639-648.

\item[]\label{rol97} Rol\'{a}n-Alvarez, E., Johannesson, K. \& Erlandson, J.
1997. The maintenance of a cline in the marine snail {\em Littorina Saxatilis}:
the role of home site advantage and hybrid fitness. {\em Evolution} {\bf 51},
1838-1847.

\item[]\label{sch98} Schluter, D. 1998 Ecological causes of speciation. 
In {\em Endless forms: species and speciation} 
(eds Howard, D. J. \& Berlocher, S. H.), pp. 114-129. 
New York: Oxford University Press.

\item[]\label{sin90} Singh, R. S. 1990 Patterns of species divergence and genetic 
theories of speciation. In {\em Population Biology. Ecological and 
Evolutionary Viewpoints} (eds W\"{o}hrmann, K. \& Jain, C. K.), 
pp. 231-265. Berlin: Springer-Verlag.

\item[]\label{sla76} Slatkin, M. 1976 The rate of spread of an advantageous allele in a 
subdivided population. In {\em Population genetics and ecology}. (eds Karlin, 
S. \& Nevo, E.), pp. 767-780. New York: Academic Press.

\item[]\label{sla81} Slatkin, M. 1981 Fixation probabilities and fixation times in a 
subdivided population. {\em Evolution} {\bf 35}, 477-488.

\item[]\label{sla82} Slatkin, M. 1982 Pleiotropy and parapatric speciation. 
{\em Evolution} {\bf 36}, 263-270.

\item[]\label{sla87} Slatkin, M. 1987 Gene flow and the geographic structure 
of natural populations. {\em Science} {\bf 236}, 787-792.

\item[]\label{tac91} Tachida, H. \& Iizuka, M. 1991 Fixation probability 
in spatially varying environments. {\em Genet. Res.} {\bf 58}, 243-251.

\item[]\label{tur95} Turner, G. F. \& Burrows, M. T. 1995. A model of
sympatric speciation by sexual selection. {\em Proc. R. Soc. Lond.} B {\bf 260},
287-292.

\item[]\label{vac98} Vacquier, V. D. 1998 Evolution of gamete recognition proteins. 
{\em Science} {\bf 281}, 1995-1998.

\item[]\label{van98} van Doorn, G. S. , Noest, A. J. \& Hogeweg, P. 1998 
Sympatric speciation and extinction driven by environment dependent sexual
selection. {\em Proc. R. Soc. Lond.} B, {\bf 265}, 1915-1919.

\item[]\label{vrb} Vrba, E. S. 1985. Environment and evolution: alternative
causes of the temporal distribution of evolutionary events. {\em South
African Journal of Science} {\bf 81} 229-236.

\item[]\label{wad84} Wade, M. J. \& McCauley, D. E. 1984 Group selection:
the interaction of local deme size and migration in the differentiation of
small populations. {\em Evolution} {\bf 38}, 1047-1058.

\item[]\label{wal82} Walsh, J. B. 1982 Rate of accumulation of reproductive 
isolation by chromosome rearrangements. {\em Amer. Natur.} {\bf 120}, 510-532.

\item[]\label{wil77} Wills, C. J. 1977 A mechanism for rapid allopatric 
speciation. {\em Amer. Natur.} {\bf 111}, 603-605.

\item[]\label{wu85} Wu, C-I. 1985 A stochastic simulation study of speciation 
by sexual selection. {\em Evolution} {\bf 39}, 66-82.

\item[]\label{wu94} Wu, C. -I. \& Palopoli, M. F.  1994
Genetics of postmating reproductive isolation in animals.
{\em Ann. Rev. Genet.} {\bf 27}, 283-308.

\end{enumerate}

\end{document}